\documentclass[aps,prl,reprint,groupedaddress]{revtex4-1}
\usepackage{graphicx}
\usepackage{color}

\setcitestyle{numbers,square}

\begin{document}


\title{Evolution of Magnetic and Orbital Properties in the Magnetically-Diluted \textit{A}-Site Spinel Cu$_{1-x}$Zn$_x$Rh$_2$O$_4$}


\author{A. V. Zakrzewski}
\author{S. Gangopadhyay}
\affiliation{Department of Physics and Seitz Materials Research Laboratory, University of Illinois at Urbana-Champaign, Urbana, Illinois 61801, USA}

\author{A. A. Aczel}
\author{S. Calder}
\author{T. J. Williams}
\affiliation{Neutron Sciences Directorate, Oak Ridge National Laboratory, Oak Ridge, Tennessee 37831, USA}

\author{G. J. MacDougall}
\affiliation{Department of Physics and Seitz Materials Research Laboratory, University of Illinois at Urbana-Champaign, Urbana, Illinois 61801, USA}
\email[gmacdoug@illinois.edu]{gmacdoug@illinois.edu}


\date{\today}

\begin{abstract}
In frustrated spinel antiferromagnets, dilution with non-magnetic ions can be a powerful strategy for probing unconventional spin states or uncovering interesting phenomena. Here, we present X-ray, neutron scattering and thermodynamic studies of the effects of magnetic dilution of the tetragonally-distorted A-site spinel antiferromagnet, CuRh$_2$O$_4$, with non-magnetic Zn$^{2+}$ ions. Our data confirm the helical spin order recently identified at low-temperatures in this material, and further demonstrate a systematic suppression of the associated N\'eel temperature with increasing site dilution towards a continuous transition with critical doping of $x_{spin} \sim 0.44$. Interestingly, this critical doping is demonstrably distinct from a second structural critical point at $x_{JT} \sim 0.6$, which is consistent with the suppression of orbital order on the A-site through a classical percolative mechanism.  This anomalously low value for  $x_{spin}$ is confirmed via multiple measurements, and is inconsistent with predictions of classical percolation theory, suggesting that the spin transition in this material is driven by an enhancement of pre-existing spin fluctuations with weak dilution.
\end{abstract}

\pacs{}

\maketitle

\section{Introduction}

The \textit{A}-site spinels are a class of materials that have received considerable attention in recent decades, both through experiment and theoretical predictions. In these materials, magnetic cations occupy the \textit{A}-sites of the spinel structure, $AB_2O_4$, and comprise a diamond sublattice. The diamond lattice is bipartite, and \textit{A}-site spinels with dominant near-neighbor exchange interactions tend to exhibit a form of N\'eel antiferromagnetism with relatively high transition temperatures\cite{Roth64, GJM2011, Tristan2005}. However, next-nearest-neighbor interactions can destabilize this state, and the associated frustration from competing interactions can lead to a variety of interesting behaviors\cite{Lee2008, Balents2010}. A prime example is the spiral spin liquid state (SSL),\cite{Bergman2007} first identified by Bergman in consideration of the $J_1$-$J_2$ model on the diamond lattice\cite{Lee2008, Savary2011}. Here, $J_1$ and $J_2$ refer to the nearest-neighbor and next-nearest neighbor exchange terms, respectively, in the Heisenberg model Hamiltonian:

\begin{equation}
H=J_1\sum\limits_{\langle i,j\rangle}\mathbf{S_\textit{i}\cdot S_\textit{j}} + J_2\sum\limits_{\langle \langle i,j\rangle \rangle}\mathbf{S_\textit{i}\cdot S_\textit{j}}
\end{equation}

In this model, mean-field calculations show that the transition temperatures for known N\'eel antiferromagnetism decrease with increasing $J_2$ and are driven to zero at the critical point $J_2/J_1$ = 0.125, above which the novel SSL is predicted to emerge at moderate temperatures. The SSL state is defined as an infinitely degenerate set of coplanar spin spirals, and signified by a continuously occupied surface of propagation vectors \textbf{\textit{q}} in reciprocal space\cite{Bergman2007}. At lowest temperatures, it is predicted that unique \textbf{\textit{q}}-vectors on the spiral surface are selected by thermal\cite{Bergman2007} or quantum\cite{Bernier2008, Lee2008} fluctuations, driving order-by-disorder transitions to static spin ordered states that depend on the exact values of $J_1$ and $J_2$ \cite{Villain80, Henley89}. These predictions have driven a concerted effort to verify SSL properties of several \textit{A}-site spinel systems\cite{Suzuki2007, Tristan2005, Krimmel2009, Zaharko2011, GJM2011, Nair2013, Fritsch2004, Chen2009, Krimmel2007, Mucksch2007, Giri2005, Buettgen2006, Kalvius2006}, and recent neutron scattering investigations have recently verified key SSL predictions in the manganese thiospinel MnSc$_2$S$_4$\cite{Gao2017}. In oxides, attention has mostly focused on the compound CoAl$_2$O$_4$, which was proposed as another SSL candidate early on due to its demonstrable level of frustration and short-ranged magnetic correlations\cite{ Krimmel2009, Bergman2007}. Rather than SSL behavior however, extensive studies have revealed an unconventional ground state in this material dominated by frozen N\'eel correlations\cite{GJM2011, GJM2016, Tristan2005, Krimmel2009, Zaharko2011, Roy2013, Nair2013}. Very recently work has open a new line of research in rhodium spinels NiRh$_2$O$_4$\cite{chamorro2018},  CoRh$_2$O$_4$\cite{Mourigal2017} and  CuRh$_2$O$_4$\cite{Mourigal2017}, which have smaller moments than in Mn or Co systems and evidence for strong spin fluctuations.

In this Article, we focus on the latter of these new \textit{A}-site spinel oxide, CuRh$_2$O$_4$. This compound is a normal spinel in which S = 1/2 Cu$^{2+}$ ions occupy the tetrahedral A-sites, while Rh$^{3+}$ ions occupy octahedral B-sites and are non-magnetic due to a large crystal field gap. (See Fig. 1.) As with most spinels, this material crystallizes in a high-temperature cubic structure (space group: $Fd\bar{3}m$), but here undergoes a Jahn-Teller distortion at 850 K\cite{Kennedy99} driven by degeneracy of Cu$^{2+}$ $e_g$ orbitals and assumes a tetragonal crystal structure at low temperatures (space group: $I4_1/amd$)\cite{Dollase97}. Bulk magnetization and specific heat measurements\cite{Endoh99} indicate the onset of antiferromagnetism with N\'{e}el temperature  $T_N = 21.9$ K, presumably associated with incommensurate helical order (propagation vector $\textbf{\textit{k}}_m$ = (0, 0, 0.79)) reported by a  very recent magnetic neutron scattering study\cite{Mourigal2017}.  Notably, this same neutron study flagged  CuRh$_2$O$_4$ as a guidepost for exotic quantum behaviour due to the significant level of spin fluctuations inferred from the strong reduction in ordered moment size\cite{Mourigal2017}.

Although the preferred method for exploring fluctuations and exotic spin physics in A-site spinels has been detailed measurements of neutron diffuse correlations\cite{Gao2017, GJM2016, Zaharko2011, Nair2013}, the lack of large single crystals of CuRh$_2$O$_4$ has eliminated that route of exploration as a viable option. In the current Article, we instead study the effect of systematic site dilution on spin, orbital and thermodynamic properties of this compound. This method has been used successfully in studying the properties of other spinel systems. In the current context, perhaps most notable are studies of CoAl$_2$O$_4$, where dilution of the magnetic sublattice with non-magnetic Zn$^{2+}$ revealed the existence of a rich phase diagram with a variety of competing phases\cite{Naka2015} and Ga$^{3+}$ substitution at the B-site gave important information about the effects of lattice expansion and site disorder\cite{Melot2009}. In CuRh$_2$O$_4$, the motivation for doping studies is buffeted by reports of superconductivity in chemically equivalent compounds CuRh$_2$S$_4$\cite{Bitoh92} and CuRh$_2$Se$_4$\cite{Hagino95}, which raises the intriguing possibility of an unconventional superconducting state nearby in the \textit{A}-site spinel phase diagram.

For these reasons, we have synthesized and carried out an extensive study of the hole doped family Cu$_{1-x}$Zn$_x$Rh$_2$O$_4$ with end members CuRh$_2$O$_4$ and ZnRh$_2$O$_4$; the latter is a \textit{p}-type semiconductor\cite{Arlett68, Mizoguchi2002, MH2010}. Magnetization was performed to explore spin properties and search for signatures of superconductivity, while the evolution of the Jahn-Teller transition was tracked using powder x-ray diffraction. The evolution of local spin order was further probed with both neutron powder diffraction and elastic neutron scattering with a triple-axis spectrometer. Our data confirm the helical ordered state in the parent material, CuRh$_2$O$_4$, recently reported by Ge \textit{et al.}, while showing that it is continuously suppressed by hole-doping with a critical value of $x_{spin} \sim$ 0.40. The tetragonal structure associated with Jahn-Teller physics is seen to persist to higher levels of doping, before reverting to a cubic structure above $x_{JT} \sim$ 0.6. No signature of superconductivity was seen at any doping, however the difference between $x_{spin}$ and $x_{JT}$, and in particular to significant suppression of the spin critical point from predictions of percolation theory, is taken as evidence for the existence of appreciable spin-fluctuations in this material family.

\begin{figure}[h]
\includegraphics[width=\linewidth]{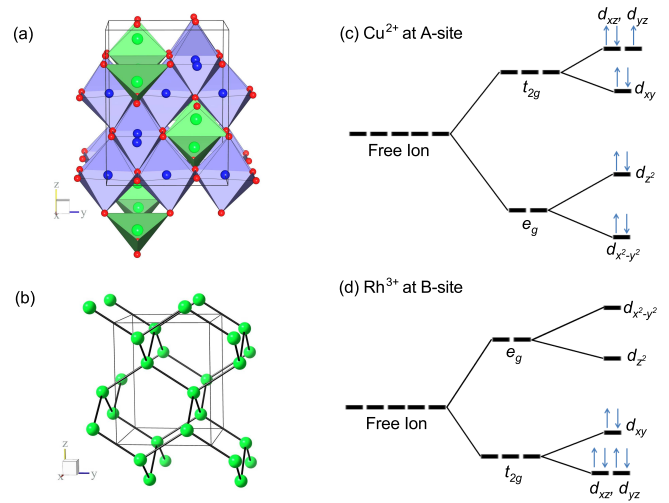}
\caption{\label{Fig1} (a) Room temperature crystal structure of tetragonally-distorted spinel CuRh$_2$O$_4$. Green spheres are Cu$^{2+}$ ions; Blue Rh$^{3+}$; Red O$^{2-}$. A-site tetrahedra and B-site octahedra are depicted with green and blue faces, respectively.
(b) Connectivity of nearest-neighbor Cu$^{2+}$ sites. (c) Crystal field splitting of d-orbital energy levels of A-sites in a tetragonally-distorted spinel with $c/a > 1$.
(d) Crystal field splitting of d-orbital energy levels of B-sites in a tetragonally-distorted spinel.}
\end{figure}

\section{Experimental Details}

Polycrystalline samples investigated in this study were synthesized via solid state reaction in the Seitz Materials Research Laboratory. Stoichiometric amounts of Rh$_2$O$_3$ (Alfa Aesar, 99.9\%), CuO (Alfa Aesar Puratronic, 99.995\%), and ZnO (Sigma-Aldrich, ACS Reagent $\geq$99\%) were weighed out, and mixed thoroughly using an ethanol slurry. The powders were subsequently pelletized and underwent solid state reaction in Pt crucibles under flowing O$_2$ gas at 1373 K for 96 h\cite{Dollase97}. Powder X-ray diffraction patterns (Cu K-$\alpha$ radiation, $\lambda$ = 1.5418 {\AA}) were measured on all powders to verify sample quality and purity, and to gain information about the structure at room temperature. Magnetization and heat capacity measurements were performed on selected samples across the entire doping axis with a Quantum Design SQUID Magnetometer (MPMS3) and a Physical Property Measurement System (PPMS), respectively.

Neutron powder diffraction experiments were carried out on several dopings of Cu$_{1-x}$Zn$_x$Rh$_2$O$_4$ using the HB2a Neutron Diffractometer\cite{Garlea2010} at Oak Ridge National Laboratory's High-Flux Isotope Reactor (HFIR). Approximately 2-3 g of powder of each composition were loaded into aluminum sample cans and mounted in a standard closed-cycle refrigerator (CCR). Due to the relatively high neutron absorption cross-section of rhodium, annular cans were used to maximize scattering intensity. $\lambda$ = 2.4111 {\AA} (Ge-113 monochromator) neutrons were used on samples expected to show magnetic order, while $\lambda$ = 1.5399 {\AA} (Ge-115 monochromator) was chosen for measurement of structural peaks at higher dopings where magnetism is suppressed. Collimations were set to open-21$'$-12$'$. In low-Zn samples where magnetism was expected, patterns were measured at $T$ = 40 K and $T$ = 4 K, well separated from $T_N$. At higher Zn dopings, diffraction patterns were measured at room temperature and $T$ = 4 K, to obtain information on structural evolution with temperature. Rietveld refinement was performed on both X-ray and neutron powder diffraction data using the \texttt{FULLPROF} suite\cite{FULLPROF93}. In addition to powder diffraction measurements, elastic scattering measurements were performed on the parent material CuRh$_2$O$_4$ using HFIR's HB1a Fixed-Incident Energy Triple-Axis Spectrometer (FIE-TAX) to further explore the helical ordered state. A $\sim$3 g sample was loaded into a standard Al can and cooled using a CCR. Fixed incident energy neutrons of $E = 14.6$ meV were used, and collimation divergences were set to 40$'$-40$'$-40$'$-80$'$. \newline

\section{Results}

\subsection{Powder X-Ray Diffraction}

\begin{figure}[h]
\includegraphics[width=\linewidth]{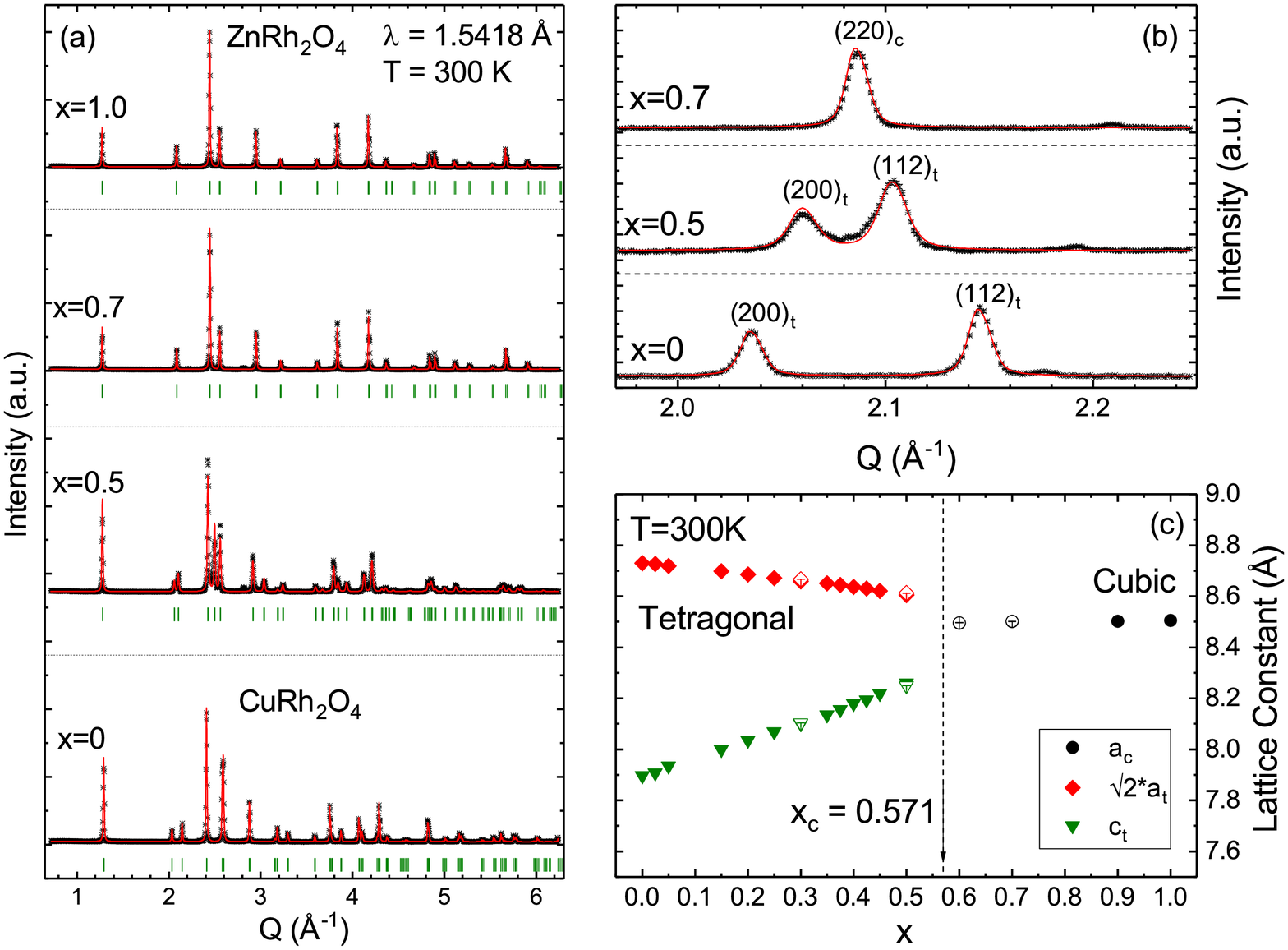}
\caption{\label{Fig2} (a) Room temperature powder XRD patterns of select dopings of Cu$_{1-x}$Zn$_x$Rh$_2$O$_4$. Crosses are data, while solid
curves represent fits described in the main text. Vertical lines are indexed reflections in  $I4_1/amd$ and  $Fd\bar{3}m$ space groups, as appropriate. (b) Doping progression of the splitting of the cubic (220) peak, which can be associated with the known Jahn-Teller transition in undoped CuRh$_2$O$_4$. (c) Lattice parameters extracted from Rietveld refinements of  XRD (closed circle) and NPD (open circle) data, revealing a cross-over from tetragonal and cubic structure near $x_{JT} \sim 0.6$. Also shown is the value of $x_c$ expected from classical percolation theory on a diamond lattice.}
\end{figure}

\begin{figure*}[tbh]
\includegraphics[width=1.5\columnwidth]{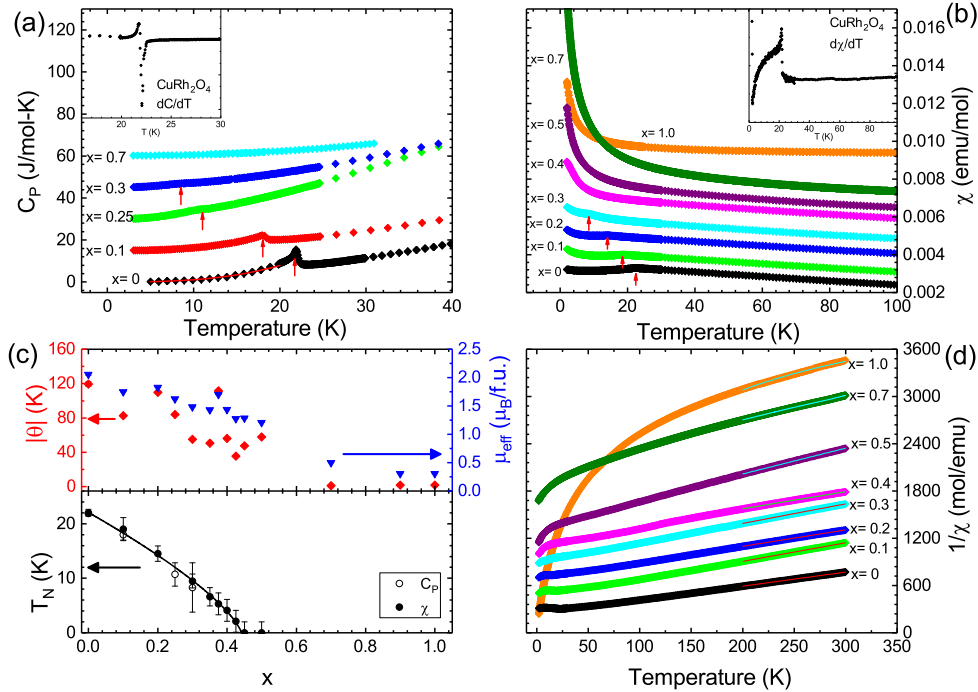}
\caption{\label{Fig3} Temperature dependence of the heat capacity (a) and magnetic susceptibility (b) for selected samples in the doping series Cu$_{1-x}$Zn$_x$RhO$_4$. Here, vertical arrows denote the position of the N\'eel temperature, defined as the location of the discontinuity of the first temperature derivative (see Insets). For clarity of the plots, curves in (a) were incrementally offset by an amount 15 J/mol-K for each successive doping, and curves in (b) were offset by an amount 0.001 emu/mol (0.003 for x=1). (c) Doping variation of $T_N$, determined from (a) and (b), along with effective moments $\mu_{eff}$, and Weiss constants $\theta$ determined from Curie-Weiss fits. The solid curve represents a fit to Eq.~\ref{eq:MF_doping}. (d) Plots of inverse susceptibility over the full measured temperature range, along with lines of best fit to Curie-Weiss form. For clarity, curves are offset by an amount 200 mol/emu (0 mol/emu for x=1).}
\end{figure*}

Fig. 2 presents a summary of results from our room temperature powder X-ray diffraction data for the Cu$_{1-x}$Zn$_x$Rh$_2$O$_4$ family, which reveal a gradual evolution from tetragonal to cubic symmetry as the system evolves from CuRh$_2$O$_4$ to ZnRh$_2$O$_4$ sides of the phase diagram. In Fig. 2(a), patterns are shown for the end members of the series and select intermediate dopings, along with solid lines showing quality of fits. ZnRh$_2$O$_4$  was confirmed to have the cubic spinel structure with space group \textit{Fd}$\bar3$\textit{m} and lattice parameter $a$ = 8.50549(6) \AA, while CuRh$_2$O$_4$ demonstrated the significant tetragonal distortion known in this material to be associated with a Jahn-Teller distortion (space group \textit{I}4$_1$/\textit{amd}, $a$ = 6.17262(6) \AA, $c$ = 7.89701(9) \AA)\cite{Endoh99}. This behavior and all measured lattice parameters are consistent with existing literature\cite{Dollase97, Kennedy99}. In addition, refinements show a level of site inversion that was within error equal to zero; though this also is consistent with previous literature\cite{Mourigal2017} and with the large difference between the atomic radii of Cu$^{2+}$ and Rh$^{3+}$, it is still a significant observation, as site inversion is a major source of disorder in spinels and contributes heavily to the magnetic properties of related materials, such as CoAl$_2$O$_4$\cite{Savary2011, Hanashima2013, Naka2015}.

At intermediate dopings, samples are seen to be single phase and evolve gradually from strongly tetragonal  (\textit{I}4$_1$/\textit{amd}) towards cubic symmetry with increasing doping, before suddenly jumping to  \textit{Fd}$\bar3$\textit{m} symmetry. There are no indications of phase co-existence in any sample measured, though x-ray data (and neutron data presented below) indicate the existence of an unknown impurity in all samples with an estimated volume of $< 5\%$ in most samples. Peaks associated with the same impurity are observed in other scattering studies of rhodium based spinels\cite{Mourigal2017}, and are more prominent in samples (even at the same doping) if they were exposed to lower oxygen levels during synthesis. We thus associate it with a separate, unidentified oxide of rhodium. Most importantly, the impurity Bragg peaks are small, have no systematic dependence on doping or temperature dependence, and do not affect the main conclusions of our work in any way. The evolution from global tetragonal to cubic symmetry in the majority phase can be seen in both the refinements of the full x-ray diffraction patterns (Fig. 2(a)) and through the inspection of the splitting of particular Bragg peaks (Fig. 2(b)). In Fig. 2(c), we show the global lattice parameters extracted from refinements from all samples investigated, which seems to imply a critical doping of $x_{JT} \sim 0.6$ for the structural transition. This critical doping is very close to the predicted percolation threshold for the diamond lattice ($x_c$ = 0.571)\cite{Scholl80}, which is consistent with the tetragonal structure being driven by the Jahn-Teller active orbitals on the copper site and then systematically removed. Also included in Fig. 2(c) are data from our neutron diffraction measurements which, as we shall see in Section III.C, show that structure is unchanged in all samples between room temperature and 4 K.

\subsection{VSM Susceptibility and Heat Capacity}

Results of heat capacity and VSM susceptibility measurements are summarized in Fig. 3. Though the Jahn-Teller transition was seen to remain out of our measurement range across the doping series, the lower temperature antiferromagnet transition at $T_N = 22$ K appears prominently in the parent material as a sharp lambda anomaly in the specific heat (Fig. 3(a)) and a downturn in the magnetization (Fig.3(b)). The magnitude of these effects decrease with moderate Zn-doping, yet the discontinuities themselves persist until the transition is seen to evolve continuously to T = 0K at $x \sim 0.40$. To quantify this observation, we separately defined the N\'eel temperature as the locations of discontinuities in the first derivatives of heat capacity and magnetization data (see insets of Fig. 3(a)and (b)), and plot the results in Fig. 3(c). The solid line in this plot represents a fit to the power law

\begin{equation}
T_N(x) = T_{N0} (1 -x/ x_{spin})^{-\phi},
\label{eq:MF_doping}
\end{equation}

which results in best fit parameters $x_{spin} = 0.441\pm 0.005$, $T_{N0} = 22.1\pm0.1K$, and $\phi = 0.73\pm0.03$.  This value for critical doping for the spin order is significantly lower than the value of $x_{JT} \sim 0.60$ inferred above from x-ray scattering data, a conclusion which is consistent with the fact that at least three of our samples how a low-T tetragonal structure but show no sign of a spin transition. This is discussed further in Section IV. The fitted value for $T_{N0}$  reinforces our observations of the parent material, CuRh$_2$O$_4$.

At higher temperatures, magnetic susceptibility data demonstrate Curie-Weiss behavior (Fig. 3(d)) across the entire doping series and were fit to the functional form:

\begin{equation}
\chi(T) = \frac{C}{T - \Theta} + \chi_0 = \frac{\mu_{eff}^2}{3(T - \Theta)} + \chi_0.
\end{equation}

Here, the temperature-independent term can be associated with orbital diamagnetism from core electrons and an additional Van-Vleck paramagnetic term dominated by the heavy Rh$^{3+}$ cations\cite{Endoh99}, and was fit to $\chi_0 = 3.5 \times 10^{-5} emu/mol$ for CuRh$_2$O$_4$ and  $chi_0 =  2.68\times 10^{-4} emu/mol$ for ZnRh$_2$O$_4$. For intermediate dopings, the $\chi_0 $ was held constant during fitting to a value determined from linearly interpolating between the end points. The appropriate fit range was determined for CuRh$_2$O$_4$ (which has the largest $T_N$) by continuously raising the lower temperature bounds until fit parameters converged asymptotically, and then held constant for other samples for consistency. The Curie-like temperature dependence in ZnRh$_2$O$_4$ is similar to what is seen in previous work\cite{Mizoguchi2002}, though still surprising given the lack of local moments in this end member compound. The magnitude of the Curie tail is small enough, however, that it can be explain by postulating a minute amount of A-B site inversion in this material, leading to a dilute level of $S$=2 $Rh^{3+}$ spins in the tetrahedral environment of the A-site. We estimate that an inversion as small as $0.4\%$ would explain the size of the current effect, which is well below our detection limits. The value of $\Theta$ for ZnRh$_2$O$_4$ is within error equal to zero, in further support of the dilute moment picture.

Fitted values $\mu_{eff}$ and $\Theta$ for the entire range of dopings are shown in Fig.3(c). Values for the $x$=0 and 1 samples are consistent with literature\cite{Endoh99, Mizoguchi2002}. Neither $\mu_{eff}$ nor $|\Theta|$ shows any signature of the $x_{spin}$ and $x_{JT}$ transitions, but both show a nearly linear decrease with $x$, as expected for a system where moments are being selectively doped out.

At low temperatures, the magnetization in all samples exhibit a small upturn with decreasing temperature, comparable to the near negligible Curie tail seen in ZnRh$_2$O$_4$. The low temperature heat capacity in the ordered state, however, is believed to reflect the density of low-lying magnetic excitations. Indeed, we observe the heat capacity of CuRh$_2$O$_4$ to vary like $C/T \propto T^{2.85}$ in the range $1.8 \textrm{K} < T < 20 $K, which is close to the $T^3$ form expected for both a system of gapless magnons in three dimensions (and also for phonons). The work of Ge \textit{et al.}\cite{Mourigal2017} describes heat capacity of their samples as varying according to

\begin{equation}
\frac{C(T)}{T} = \gamma + \alpha T^2,
\end{equation}

where they postulate that the linear term of the heat capacity is associated with a small amount of glassiness. For comparison to this previous work, we fit our data to the same function over the same range and include the result as a solid line in Fig. 3(a). The fitted values of $\alpha = 9.5(2)\times 10^{-4}$  J mol$^{-1}$ K$^{-4}$ and  $\gamma = 1.4(1)\times 10^{-2}$  J mol$^{-1}$ K$^{-2}$, are in line with previous work\cite{Mourigal2017}. Whereas Ge \textit{et al.} ascribe a significant role for phonons in the $T^3$ behavior in the low temperature specific heat, we observe that this term almost completely collapses with doping as one leaves the spin-ordered state; the heat capacity of samples with $x$ = 0.425 and 0.7 is described by a nearly linear temperature dependence of similar magnitude to that in undoped CuRh$_2$O$_4$. As we do not expect such a significant change in the phonon density across the doping series, we thus associate the cubic heat capacity in our systems with low-energy antiferromagnetic magnons. Its disappearance near $x \sim$ 0.4 is further support of the existence of a quantum critical point in the region associated with the helical spin order.

\subsection{Neutron Scattering}

\begin{figure*}[tbh]
\includegraphics[width=1.5\columnwidth]{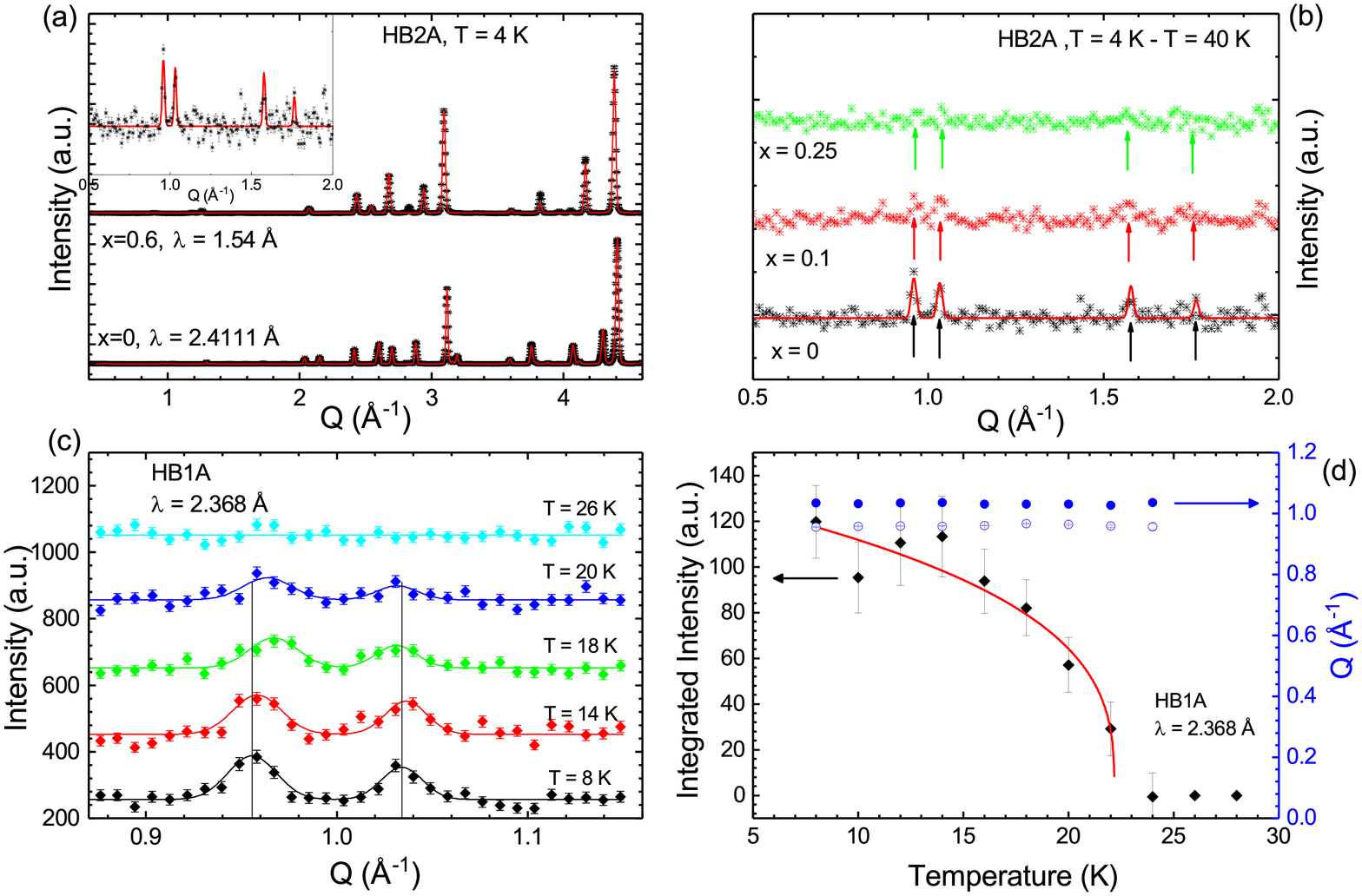}
\caption{\label{Fig4} (a) Neutron powder diffraction (NPD) patterns for $x$=0 and $x$=0.6 samples at $T$=4 K, taken with the HB2a instrument.  Inset: Low-q neutron powder diffraction data for CuRh$_2$O$_4$, denoting the location of magnetic Bragg peaks at $Q_1 = 0.97$ {\AA}$^{-1}$ and $Q_2 = 1.04$ {\AA}$^{-1}$. The solid line is the result of fits to the helical model of Ref.~\onlinecite{Mourigal2017}. (b) $T = 4$K - $T$ = 40 K difference plots in the region around the primary magnetic Bragg peaks for samples with three different levels of Zn-doping, $x$. Arrows denote the position of predicted magnetic peaks. (c) Scattering about low-q magnetic Bragg peaks in CuRh$_2$O$_4$, as measured with the HB1a triple-axis spectrometer. Data are presented for select temperatures below and above the N\'eel temperature, determined below. Solid lines are fits to a Gaussian peak shape. (d) Temperature evolution of parameters from fits in (c). Included are plots of total intensity (diamond), and peak position (circles).}
\end{figure*}

Neutron studies were carried out at the HFIR in two steps: first measuring the global magnetic and lattice structure using the HB2a powder diffractometer across the doping phase diagram, followed up by detailed temperature measurements of spin correlations in the parent compound CuRh$_2$O$_4$ using the HB1a triple-axis spectrometer. The two data sets on CuRh$_2$O$_4$ are consistent with each other, and with the helical spin ordering model recently proposed by Ge \textit{et al.}\cite{Mourigal2017} High temperature diffraction measurements reflected the same lattice structure as determined from x-ray measurements in Fig. 2. At low temperatures, we see the emergence of several incommensurate magnetic Bragg peaks at low-$Q$, with the most prominent located at $Q_1 = 0.97$ {\AA}$^{-1}$ and $Q_2 = 1.04$ {\AA}$^{-1}$. This can be seen in Fig. 4(a), which shows the neutron diffraction data at T = 4K, along with solid lines representing fits to the helical model. Fits describe the data well and give a value of $\vec{k} = (0, 0, 0.8014(7))$ for the ordering wavevector and $\mu = 0.20(5) \mu_B$ for the ordered moment size. This is a significantly smaller value than seen in the one other neutron scattering study on this compound\cite{Mourigal2017}, implying an even stronger role for quantum fluctuations than previously believed. The disagreement on moment size between the current study and Ref.~\onlinecite{Mourigal2017} stands in stark contrast to the close agreement between the two studies in all other aspects of the spin order and material properties. On this point, we note that the current study was performed using a neutron powder diffractometer specifically designed to determine magnetic structures, rather than using a neutron spectrometer\cite{Mourigal2017},  which has lower precision in the elastic channel. However, we concede that  the weak neutron scattering signal due to the strongly renormalized moment and the complicated incommensurate structure makes reliable refinements of absolute moment sizes in this compound particularly difficult.

In Fig. 4(c), we present data from HB1a showing the evolution of the scattering about the positions of the two most intense magnetic Bragg peaks at several temperatures below and above $T_N$. Solid lines represent fits to Gaussian lineshapes, and the fitted intensities of the two peaks were combined to lower relative errors and the result is plotted in Fig. 4(d) along with the peak positions. Both peak positions and widths appear to be temperature independent within the resolution of the instrument, but the intensity varies in the expected way for a magnetic order parameter. We do not have the appropriate density of points near the transition to comment on critical behaviour, however in order to quantify the temperature evolution, we chose to fit to the empirical model:

\begin{equation}
I(t) = I_0 ( 1 - \frac{T}{T_N} ) ^{p},
\end{equation}

which gave a value for N\'eel temperature of $T_N = 22.2(3)$ K. This ordering temperature is again consistent with above results from magnetization and heat capacity and with literature\cite{Endoh99}. The fitted value of $p$ was 0.16(4) though, as stated above, it should not be associated with a critical exponent.

Variation with doping was primarily explored with HB2a. In Fig. 4(b), we show low-Q neutron diffraction results from members of the Cu$_{1-x}$Zn$_x$Rh$_2$O$_4$ series with $x = 0$, $0.10$ and $0.25$. In order to emphasize the variation of $Q_1$ and $Q_2$ with doping, we plot the difference I($T$=4K) - I($T$=40K) to eliminate background scattering. Evidence for the same magnetic Bragg peaks was apparent in all samples (denoted by arrows in Fig 4(b)), though the scattering intensity decreases dramatically with Zn-doping (dropping to 60$\%$ and 45$\%$ of original values by $x$ = 0.10 and 0.25, respectively). This drop in intensity is consistent with the rapidly decreasing ordered moment inferred from magnetization results in Fig.~2. As with temperature, we observed no variation in peak position or peak width with doping, $x$. The peak seen near $Q\sim 1.95\AA^{-1}$ is seen in all samples, exists outside the helical ordering model and is evidently unaffected by the change in ordered moment size with doping. We thus attribute it to the same impurity phase identified in Fig. 2.

For higher values of $x$, ordered moments are too small to be apparent in the NPD data, though structure was examined at high and low temperature. As determined with room temperature x-ray diffraction, our NPD results show that the lattice structure changes from tetragonal \textit{I}4$_1$/\textit{amd} symmetry to cubic \textit{Fd}$\bar3$\textit{m} symmetry between dopings $x$ = 0.50 and $x$ = 0.60. Supplementing the x-ray results, NPD further shows that the lattice symmetry is the same between T = 300K and T = 4K for every sample measured, consistent with a nearly vertical boundary between the two structures in the $x$-$T$ phase diagram. Fig. 4(a) shows a representative NPD pattern for a cubic sample at high doping ($x = 0.6$), while lattice parameters extracted from our refinements of neutron data are included in Fig. 2(c).

\section{Discussion and Conclusions}

The collective experimental results of previous sections were combined to create the temperature-doping phase diagram for Cu$_{1-x}$Zn$_x$Rh$_2$O$_4$ shown in Fig.~5, which constitutes the main conclusion of this study. The boundary between orbitally ordered and helimagnet phases is most precisely determined, having corroborating signatures in magnetization, heat capacity and (for $x$=0) neutron scattering data. Below this line, our data confirms that the spins exhibit the incommensurate signatures of the helimagnetic state explored in detail in Ref.~\onlinecite{Mourigal2017}.  The phase boundary between cubic and tetragonal phases is also clearly defined at T=0K, with both x-ray and neutron diffraction showing a clean transition near $x_{JT} \sim 0.60$ and no visible phase co-existence at any doping range. As the tetragonal phase is associated with the flattening of CuO$_4$ tetrahedra in the $c$-axis direction, it is reasonable to associate this symmetry lowering structural transition with Jahn-Teller physics, as has been done in the past\cite{Kennedy99, Blasse61, Bertaut59}. At $x$=0, this transition has been seen with temperature in previous studies, with a $T_c$ = 850K\cite{Kennedy99}. This temperature is out of the range of current experiments, and in fact we did not observe any structural transition with temperature in any of our samples in the range 4K $< T <$ 300K. Our diffraction data did reveal a doping evolution of the magnitude of the tetragonal distortion, however, which we loosely associate with the strength of the Jahn-Teller interaction. For illustrative purposes then, we include a line in Fig. 5 separating high-temperature paramagnetic and orbitally ordered phases which evolves with doping in a way that parallels the magnitude of the structural distortion.

The resulting picture is one in which the orbital ordered and helimagnetic states of the quantum magnetic material CuRh$_2$O$_4$ disappears with doping in \textit{two} steps, with the spin ordered state disappearing first and orbital order at a measurably higher level of dilution. On its face, this is a puzzling situation, as both transitions are being driven by the same Cu$^{2+}$ cations and the relevant (identical) sublattice is being diluted at the same rate. On closer inspection however, it seems to imply that the two doping transitions are being driven through different mechanisms. The location of $x_{JT}$ is nearly identical to the predicted percolation threshold for a diamond lattice\cite{Scholl80}, which is also where Naka \textit{et al.} observed the disappearance of spin order signatures in site-diluted A-site spinel system Co$_{1-x}$Zn$_x$Al$_2$O$_4$\cite{Naka2015}. The spin transition in our system, on the other hand, disappears much more quickly, which would seem to indicate that the spin transition is being assisted by the doping-enhancement of significant spin fluctuations, rather than being driven by purely percolative effects. The existence of a high density of spin-fluctuations is consistent with the large renormalization of ordered moment sizes inferred from the current neutron scattering data, and that in Ref.~\onlinecite{Mourigal2017}. The enhancement of these fluctuations with dilution stands in contrast to previous studies on Co$_{1-x}$Zn$_x$Al$_2$O$_4$\cite{Naka2015}, where ordering temperatures are seen to increase with moderate doping, and thus presumably the density of fluctuations is decreasing. This may be associated with the different connectivity of atoms in cubic and tetragonally distorted diamond lattices, as discussed below.

The existence of spin fluctuations in CuRh$_2$O$_4$ might be thought of as a natural consequence of the $S = 1/2$ moment associated with the $d^9$ configuration of Cu$^{2+}$ cations. In the current class of compounds however, it takes on special significance. As laid out in the Introduction section, one of the major factors driving research into the larger class of \textit{A}-site spinels is the possibility of quantum ground states, up to and including the SSL phase, which stems from exchange competition and is associated with a near degeneracy of an infinite number of spiral spin states\cite{Bergman2007, Lee2008, Gao2017}.  In cubic spinels, competition is between $J_1$ and $J_2$, and since each atom on the diamond lattice has four nearest-neighbors and twelve next-nearest-neighbors, one would expect the primary effect of dilution would be to decrease the ratio $\frac{J_2}{J_1}$. For the material CoAl$_2$O$_4$, with $\frac{J_2}{J_1} \lesssim \frac{1}{8}$, this would result in an increase in N\'eel temperature, as is observed\cite{Naka2015}. In CuRh$_2$O$_4$ however, the picture is modified significantly by the presence of the large Jahn-Teller distortion, which breaks the equivalency between next-nearest-neighbor bonds out-of ($J_2$) and in ($J_2'$) the $ab$-plane. Indeed, the mean field calculations of Ge \textit{et al.}\cite{Mourigal2017} show that the incommensurate ordered states with propagation vectors $\vec{k} = (0, 0, \xi)$ are only stable in regions when $\frac{J_2}{J_1}$ is relatively strong and $\frac{J_2'}{J_1}$ relatively weak. In doping Cu$_{1-x}$Zn$_x$Rh$_2$O$_4$ from $x$=0 to  $x$=0.30, our data reveal that out-of-plane next-nearest neighbor distances increase by 0.5$\%$  and in-plane distances decrease by 0.8$\%$, implying an associated drop in $J_2$ and increase in $J_2'$. As it has already been determined that CuRh$_2$O$_4$ lies on the cusp of the stability region for the $\vec{k} = (0, 0, \xi)$ phase\cite{Mourigal2017}, fluctuations in the current study may be interpreted as competition from the adjacent $\vec{k} = (\xi, 0, 0)$ phase predicted for materials with only slightly larger $\frac{J_2'}{J_2}$. This reentrant frustration with doping in turn implies the existence of a high density of low-lying competing states and strongly motivates single crystal work.

On a more speculative note, we again point to the existence of superconducting phases in closely related materials CuRh$_2$S$_4$\cite{Bitoh92} and CuRh$_2$Se$_4$\cite{Hagino95}, whose properties have not yet been studied in detail. We observed no signatures of superconductivity in the current material family, and indeed the samples remained insulating across the phase diagram. Nonetheless, the current evidence for enhanced spin fluctuations in CuRh$_2$O$_4$ with hole-doping further provides motivation to study this material with pressure or doping on the Rh-site in order to induce metallicity and, conceivably, an unconventional superconducting state.

\begin{figure}[h]
\includegraphics[width=\linewidth]{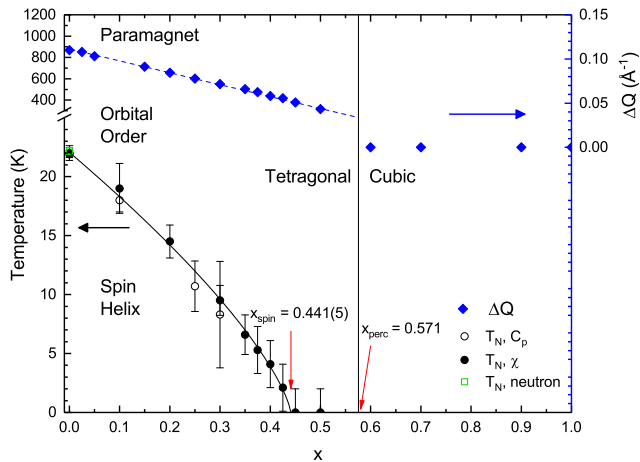}
\caption{\label{Fig5} Composite phase diagram for Cu$_{1-x}$Zn$_x$Rh$_2$O$_4$, constructed from x-ray diffraction, neutron scattering, specific heat, and DC susceptibility measurements laid out in the current article. Open (filled) circles correspond to $T_N$ obtained from specific heat (magnetization). The open square corresponds to $T_N$ extracted from triple-axis data (left-axis). Diamonds correspond to peak splitting of cubic (220) into tetragonal (200) and (112) structural Bragg reflections determined from x-ray scattering measurements at $T = 300$ K (right axis). A dashed line is drawn to represent the Jahn-Teller transition, under assumptions laid out in the main text. The critical doping for complete suppression of the helimagnetic transition, $x_{spin}$ is indicated by an arrow, alongside the predicted percolation threshold, $x_{perc}$, which explains the suppression of orbital order.}
\end{figure}

\begin{acknowledgments}
This work was sponsored by the National Science Foundation, under Grant No. DMR-1455264 (G.J.M., A.V.Z, and S.G.). Synthesis, x-ray and thermodynamic measurements was carried out in the Frederick Seitz Materials Research Laboratory Central Research Facilities, University of Illinois. Neutron scattering research at Oak Ridge National Laboratory's High Flux Isotope Reactor was sponsored by the Scientific User Facilities Division, Office of Basic Energy Sciences, U.S. Department of Energy.
\end{acknowledgments}


\begin{thebibliography}{42}%
\makeatletter
\providecommand \@ifxundefined [1]{%
 \@ifx{#1\undefined}
}%
\providecommand \@ifnum [1]{%
 \ifnum #1\expandafter \@firstoftwo
 \else \expandafter \@secondoftwo
 \fi
}%
\providecommand \@ifx [1]{%
 \ifx #1\expandafter \@firstoftwo
 \else \expandafter \@secondoftwo
 \fi
}%
\providecommand \natexlab [1]{#1}%
\providecommand \enquote  [1]{``#1''}%
\providecommand \bibnamefont  [1]{#1}%
\providecommand \bibfnamefont [1]{#1}%
\providecommand \citenamefont [1]{#1}%
\providecommand \href@noop [0]{\@secondoftwo}%
\providecommand \href [0]{\begingroup \@sanitize@url \@href}%
\providecommand \@href[1]{\@@startlink{#1}\@@href}%
\providecommand \@@href[1]{\endgroup#1\@@endlink}%
\providecommand \@sanitize@url [0]{\catcode `\\12\catcode `\$12\catcode
  `\&12\catcode `\#12\catcode `\^12\catcode `\_12\catcode `\%12\relax}%
\providecommand \@@startlink[1]{}%
\providecommand \@@endlink[0]{}%
\providecommand \url  [0]{\begingroup\@sanitize@url \@url }%
\providecommand \@url [1]{\endgroup\@href {#1}{\urlprefix }}%
\providecommand \urlprefix  [0]{URL }%
\providecommand \Eprint [0]{\href }%
\providecommand \doibase [0]{http://dx.doi.org/}%
\providecommand \selectlanguage [0]{\@gobble}%
\providecommand \bibinfo  [0]{\@secondoftwo}%
\providecommand \bibfield  [0]{\@secondoftwo}%
\providecommand \translation [1]{[#1]}%
\providecommand \BibitemOpen [0]{}%
\providecommand \bibitemStop [0]{}%
\providecommand \bibitemNoStop [0]{.\EOS\space}%
\providecommand \EOS [0]{\spacefactor3000\relax}%
\providecommand \BibitemShut  [1]{\csname bibitem#1\endcsname}%
\let\auto@bib@innerbib\@empty
\bibitem [{\citenamefont {Roth}(1964)}]{Roth64}%
  \BibitemOpen
  \bibfield  {author} {\bibinfo {author} {\bibfnamefont {W.~L.}\ \bibnamefont
  {Roth}},\ }\href@noop {} {\bibfield  {journal} {\bibinfo  {journal} {J. Phys.
  France}\ }\textbf {\bibinfo {volume} {25}},\ \bibinfo {pages} {507} (\bibinfo
  {year} {1964})}\BibitemShut {NoStop}%
\bibitem [{\citenamefont {MacDougall}\ \emph {et~al.}(2011)\citenamefont
  {MacDougall}, \citenamefont {Gout}, \citenamefont {Zarestky}, \citenamefont
  {Ehlers}, \citenamefont {Podlesnyak}, \citenamefont {MacGuire}, \citenamefont
  {Mandrus},\ and\ \citenamefont {Nagler}}]{GJM2011}%
  \BibitemOpen
  \bibfield  {author} {\bibinfo {author} {\bibfnamefont {G.~J.}\ \bibnamefont
  {MacDougall}}, \bibinfo {author} {\bibfnamefont {D.}~\bibnamefont {Gout}},
  \bibinfo {author} {\bibfnamefont {J.~L.}\ \bibnamefont {Zarestky}}, \bibinfo
  {author} {\bibfnamefont {G.}~\bibnamefont {Ehlers}}, \bibinfo {author}
  {\bibfnamefont {A.}~\bibnamefont {Podlesnyak}}, \bibinfo {author}
  {\bibfnamefont {M.~A.}\ \bibnamefont {MacGuire}}, \bibinfo {author}
  {\bibfnamefont {D.}~\bibnamefont {Mandrus}}, \ and\ \bibinfo {author}
  {\bibfnamefont {S.~E.}\ \bibnamefont {Nagler}},\ }\href@noop {} {\bibfield
  {journal} {\bibinfo  {journal} {Proc. Natl. Acad. Sci.}\ }\textbf {\bibinfo
  {volume} {108}},\ \bibinfo {pages} {15693} (\bibinfo {year}
  {2011})}\BibitemShut {NoStop}%
\bibitem [{\citenamefont {Tristan}\ \emph {et~al.}(2005)\citenamefont
  {Tristan}, \citenamefont {Hemberger}, \citenamefont {Krimmel}, \citenamefont
  {von Nidda}, \citenamefont {Tsurkan},\ and\ \citenamefont
  {Loidl}}]{Tristan2005}%
  \BibitemOpen
  \bibfield  {author} {\bibinfo {author} {\bibfnamefont {N.}~\bibnamefont
  {Tristan}}, \bibinfo {author} {\bibfnamefont {J.}~\bibnamefont {Hemberger}},
  \bibinfo {author} {\bibfnamefont {A.}~\bibnamefont {Krimmel}}, \bibinfo
  {author} {\bibfnamefont {H.-A.~K.}\ \bibnamefont {von Nidda}}, \bibinfo
  {author} {\bibfnamefont {V.}~\bibnamefont {Tsurkan}}, \ and\ \bibinfo
  {author} {\bibfnamefont {A.}~\bibnamefont {Loidl}},\ }\href@noop {}
  {\bibfield  {journal} {\bibinfo  {journal} {Phys. Rev. B}\ }\textbf {\bibinfo
  {volume} {72}},\ \bibinfo {pages} {174404} (\bibinfo {year}
  {2005})}\BibitemShut {NoStop}%
\bibitem [{\citenamefont {Lee}\ and\ \citenamefont {Balents}(2008)}]{Lee2008}%
  \BibitemOpen
  \bibfield  {author} {\bibinfo {author} {\bibfnamefont {S.}~\bibnamefont
  {Lee}}\ and\ \bibinfo {author} {\bibfnamefont {L.}~\bibnamefont {Balents}},\
  }\href@noop {} {\bibfield  {journal} {\bibinfo  {journal} {Phys. Rev. B}\
  }\textbf {\bibinfo {volume} {78}},\ \bibinfo {pages} {144417} (\bibinfo
  {year} {2008})}\BibitemShut {NoStop}%
\bibitem [{\citenamefont {Balents}(2010)}]{Balents2010}%
  \BibitemOpen
  \bibfield  {author} {\bibinfo {author} {\bibfnamefont {L.}~\bibnamefont
  {Balents}},\ }\href@noop {} {\bibfield  {journal} {\bibinfo  {journal}
  {Nature}\ }\textbf {\bibinfo {volume} {464}},\ \bibinfo {pages} {199}
  (\bibinfo {year} {2010})}\BibitemShut {NoStop}%
\bibitem [{\citenamefont {Bergman}\ \emph {et~al.}(2007)\citenamefont
  {Bergman}, \citenamefont {Alicea}, \citenamefont {Gull}, \citenamefont
  {Trebst},\ and\ \citenamefont {Balents}}]{Bergman2007}%
  \BibitemOpen
  \bibfield  {author} {\bibinfo {author} {\bibfnamefont {D.}~\bibnamefont
  {Bergman}}, \bibinfo {author} {\bibfnamefont {J.}~\bibnamefont {Alicea}},
  \bibinfo {author} {\bibfnamefont {E.}~\bibnamefont {Gull}}, \bibinfo {author}
  {\bibfnamefont {S.}~\bibnamefont {Trebst}}, \ and\ \bibinfo {author}
  {\bibfnamefont {L.}~\bibnamefont {Balents}},\ }\href@noop {} {\bibfield
  {journal} {\bibinfo  {journal} {Nat. Phys.}\ }\textbf {\bibinfo {volume}
  {3}},\ \bibinfo {pages} {487} (\bibinfo {year} {2007})}\BibitemShut {NoStop}%
\bibitem [{\citenamefont {Savary}\ \emph {et~al.}(2011)\citenamefont {Savary},
  \citenamefont {Gull}, \citenamefont {Trebst}, \citenamefont {Alicea},
  \citenamefont {Bergman},\ and\ \citenamefont {Balents}}]{Savary2011}%
  \BibitemOpen
  \bibfield  {author} {\bibinfo {author} {\bibfnamefont {L.}~\bibnamefont
  {Savary}}, \bibinfo {author} {\bibfnamefont {E.}~\bibnamefont {Gull}},
  \bibinfo {author} {\bibfnamefont {S.}~\bibnamefont {Trebst}}, \bibinfo
  {author} {\bibfnamefont {J.}~\bibnamefont {Alicea}}, \bibinfo {author}
  {\bibfnamefont {D.}~\bibnamefont {Bergman}}, \ and\ \bibinfo {author}
  {\bibfnamefont {L.}~\bibnamefont {Balents}},\ }\href@noop {} {\bibfield
  {journal} {\bibinfo  {journal} {Phys. Rev. B}\ }\textbf {\bibinfo {volume}
  {84}},\ \bibinfo {pages} {064438} (\bibinfo {year} {2011})}\BibitemShut
  {NoStop}%
\bibitem [{\citenamefont {Bernier}\ \emph {et~al.}(2008)\citenamefont
  {Bernier}, \citenamefont {Lawler},\ and\ \citenamefont {Kim}}]{Bernier2008}%
  \BibitemOpen
  \bibfield  {author} {\bibinfo {author} {\bibfnamefont {J.-S.}\ \bibnamefont
  {Bernier}}, \bibinfo {author} {\bibfnamefont {M.~J.}\ \bibnamefont {Lawler}},
  \ and\ \bibinfo {author} {\bibfnamefont {Y.~B.}\ \bibnamefont {Kim}},\
  }\href@noop {} {\bibfield  {journal} {\bibinfo  {journal} {Phys. Rev. Lett.}\
  }\textbf {\bibinfo {volume} {101}},\ \bibinfo {pages} {047201} (\bibinfo
  {year} {2008})}\BibitemShut {NoStop}%
\bibitem [{\citenamefont {Villain}\ \emph {et~al.}(1980)\citenamefont
  {Villain}, \citenamefont {Bidaux}, \citenamefont {Carton},\ and\
  \citenamefont {Conte}}]{Villain80}%
  \BibitemOpen
  \bibfield  {author} {\bibinfo {author} {\bibfnamefont {J.}~\bibnamefont
  {Villain}}, \bibinfo {author} {\bibfnamefont {R.}~\bibnamefont {Bidaux}},
  \bibinfo {author} {\bibfnamefont {J.-P.}\ \bibnamefont {Carton}}, \ and\
  \bibinfo {author} {\bibfnamefont {R.}~\bibnamefont {Conte}},\ }\href@noop {}
  {\bibfield  {journal} {\bibinfo  {journal} {J. Physique}\ }\textbf {\bibinfo
  {volume} {41}},\ \bibinfo {pages} {1263} (\bibinfo {year}
  {1980})}\BibitemShut {NoStop}%
\bibitem [{\citenamefont {Henley}(1989)}]{Henley89}%
  \BibitemOpen
  \bibfield  {author} {\bibinfo {author} {\bibfnamefont {C.~L.}\ \bibnamefont
  {Henley}},\ }\href@noop {} {\bibfield  {journal} {\bibinfo  {journal} {Phys.
  Rev. Lett.}\ }\textbf {\bibinfo {volume} {62}},\ \bibinfo {pages} {2056}
  (\bibinfo {year} {1989})}\BibitemShut {NoStop}%
\bibitem [{\citenamefont {Suzuki}\ \emph {et~al.}(2007)\citenamefont {Suzuki},
  \citenamefont {Nagai}, \citenamefont {Nohara},\ and\ \citenamefont
  {Takagi}}]{Suzuki2007}%
  \BibitemOpen
  \bibfield  {author} {\bibinfo {author} {\bibfnamefont {T.}~\bibnamefont
  {Suzuki}}, \bibinfo {author} {\bibfnamefont {H.}~\bibnamefont {Nagai}},
  \bibinfo {author} {\bibfnamefont {M.}~\bibnamefont {Nohara}}, \ and\ \bibinfo
  {author} {\bibfnamefont {H.}~\bibnamefont {Takagi}},\ }\href@noop {}
  {\bibfield  {journal} {\bibinfo  {journal} {J. Phys. Condens. Matter}\
  }\textbf {\bibinfo {volume} {19}},\ \bibinfo {pages} {145265} (\bibinfo
  {year} {2007})}\BibitemShut {NoStop}%
\bibitem [{\citenamefont {Krimmel}\ \emph {et~al.}(2009)\citenamefont
  {Krimmel}, \citenamefont {Mutka}, \citenamefont {Koza}, \citenamefont
  {Tsurkan},\ and\ \citenamefont {Loidl}}]{Krimmel2009}%
  \BibitemOpen
  \bibfield  {author} {\bibinfo {author} {\bibfnamefont {A.}~\bibnamefont
  {Krimmel}}, \bibinfo {author} {\bibfnamefont {H.}~\bibnamefont {Mutka}},
  \bibinfo {author} {\bibfnamefont {M.~M.}\ \bibnamefont {Koza}}, \bibinfo
  {author} {\bibfnamefont {V.}~\bibnamefont {Tsurkan}}, \ and\ \bibinfo
  {author} {\bibfnamefont {A.}~\bibnamefont {Loidl}},\ }\href@noop {}
  {\bibfield  {journal} {\bibinfo  {journal} {Phys. Rev. B}\ }\textbf {\bibinfo
  {volume} {79}},\ \bibinfo {pages} {134406} (\bibinfo {year}
  {2009})}\BibitemShut {NoStop}%
\bibitem [{\citenamefont {Zaharko}\ \emph {et~al.}(2011)\citenamefont
  {Zaharko}, \citenamefont {Christensen}, \citenamefont {Cervellino},
  \citenamefont {Tsurkan}, \citenamefont {Maljuk}, \citenamefont {Stuhr},
  \citenamefont {Niedermayer},\ and\ \citenamefont {Loidl}}]{Zaharko2011}%
  \BibitemOpen
  \bibfield  {author} {\bibinfo {author} {\bibfnamefont {O.}~\bibnamefont
  {Zaharko}}, \bibinfo {author} {\bibfnamefont {N.~B.}\ \bibnamefont
  {Christensen}}, \bibinfo {author} {\bibfnamefont {A.}~\bibnamefont
  {Cervellino}}, \bibinfo {author} {\bibfnamefont {V.}~\bibnamefont {Tsurkan}},
  \bibinfo {author} {\bibfnamefont {A.}~\bibnamefont {Maljuk}}, \bibinfo
  {author} {\bibfnamefont {U.}~\bibnamefont {Stuhr}}, \bibinfo {author}
  {\bibfnamefont {C.}~\bibnamefont {Niedermayer}}, \ and\ \bibinfo {author}
  {\bibfnamefont {A.}~\bibnamefont {Loidl}},\ }\href@noop {} {\bibfield
  {journal} {\bibinfo  {journal} {Phys. Rev. B}\ }\textbf {\bibinfo {volume}
  {84}},\ \bibinfo {pages} {094403} (\bibinfo {year} {2011})}\BibitemShut
  {NoStop}%
\bibitem [{\citenamefont {Nair}\ \emph {et~al.}(2013)\citenamefont {Nair},
  \citenamefont {Fu}, \citenamefont {Voigt}, \citenamefont {Su},\ and\
  \citenamefont {Br{\"u}ckel}}]{Nair2013}%
  \BibitemOpen
  \bibfield  {author} {\bibinfo {author} {\bibfnamefont {H.~S.}\ \bibnamefont
  {Nair}}, \bibinfo {author} {\bibfnamefont {Z.}~\bibnamefont {Fu}}, \bibinfo
  {author} {\bibfnamefont {J.}~\bibnamefont {Voigt}}, \bibinfo {author}
  {\bibfnamefont {Y.}~\bibnamefont {Su}}, \ and\ \bibinfo {author}
  {\bibfnamefont {T.}~\bibnamefont {Br{\"u}ckel}},\ }\href@noop {} {\bibfield
  {journal} {\bibinfo  {journal} {Phys. Rev. B}\ }\textbf {\bibinfo {volume}
  {89}},\ \bibinfo {pages} {174431} (\bibinfo {year} {2013})}\BibitemShut
  {NoStop}%
\bibitem [{\citenamefont {Fritsch}\ \emph {et~al.}(2004)\citenamefont
  {Fritsch}, \citenamefont {Hemberger}, \citenamefont {B{\"u}ttgen},
  \citenamefont {Scheidt}, \citenamefont {von Nidda}, \citenamefont {Loidl},\
  and\ \citenamefont {Tsurkan}}]{Fritsch2004}%
  \BibitemOpen
  \bibfield  {author} {\bibinfo {author} {\bibfnamefont {V.}~\bibnamefont
  {Fritsch}}, \bibinfo {author} {\bibfnamefont {J.}~\bibnamefont {Hemberger}},
  \bibinfo {author} {\bibfnamefont {N.}~\bibnamefont {B{\"u}ttgen}}, \bibinfo
  {author} {\bibfnamefont {E.-W.}\ \bibnamefont {Scheidt}}, \bibinfo {author}
  {\bibfnamefont {H.-A.~K.}\ \bibnamefont {von Nidda}}, \bibinfo {author}
  {\bibfnamefont {A.}~\bibnamefont {Loidl}}, \ and\ \bibinfo {author}
  {\bibfnamefont {V.}~\bibnamefont {Tsurkan}},\ }\href@noop {} {\bibfield
  {journal} {\bibinfo  {journal} {Phys. Rev. Lett.}\ }\textbf {\bibinfo
  {volume} {92}},\ \bibinfo {pages} {116401} (\bibinfo {year}
  {2004})}\BibitemShut {NoStop}%
\bibitem [{\citenamefont {Chen}\ \emph {et~al.}(2009)\citenamefont {Chen},
  \citenamefont {Balents},\ and\ \citenamefont {Schnyder}}]{Chen2009}%
  \BibitemOpen
  \bibfield  {author} {\bibinfo {author} {\bibfnamefont {G.}~\bibnamefont
  {Chen}}, \bibinfo {author} {\bibfnamefont {L.}~\bibnamefont {Balents}}, \
  and\ \bibinfo {author} {\bibfnamefont {A.~P.}\ \bibnamefont {Schnyder}},\
  }\href@noop {} {\bibfield  {journal} {\bibinfo  {journal} {Phys. Rev. Lett.}\
  }\textbf {\bibinfo {volume} {102}},\ \bibinfo {pages} {096406} (\bibinfo
  {year} {2009})}\BibitemShut {NoStop}%
\bibitem [{\citenamefont {Krimmel}\ \emph {et~al.}(2006)\citenamefont
  {Krimmel}, \citenamefont {M{\"u}cksch}, \citenamefont {Tsurkan},\ and\
  \citenamefont {Loidl}}]{Krimmel2007}%
  \BibitemOpen
  \bibfield  {author} {\bibinfo {author} {\bibfnamefont {A.}~\bibnamefont
  {Krimmel}}, \bibinfo {author} {\bibfnamefont {M.}~\bibnamefont
  {M{\"u}cksch}}, \bibinfo {author} {\bibfnamefont {V.}~\bibnamefont
  {Tsurkan}}, \ and\ \bibinfo {author} {\bibfnamefont {A.}~\bibnamefont
  {Loidl}},\ }\href@noop {} {\bibfield  {journal} {\bibinfo  {journal} {Phys.
  Rev. B}\ }\textbf {\bibinfo {volume} {73}},\ \bibinfo {pages} {014413}
  (\bibinfo {year} {2006})}\BibitemShut {NoStop}%
\bibitem [{\citenamefont {M{\"u}cksch}\ \emph {et~al.}(2007)\citenamefont
  {M{\"u}cksch}, \citenamefont {Tsurkan}, \citenamefont {Krimmel},
  \citenamefont {Horn},\ and\ \citenamefont {Loidl}}]{Mucksch2007}%
  \BibitemOpen
  \bibfield  {author} {\bibinfo {author} {\bibfnamefont {M.}~\bibnamefont
  {M{\"u}cksch}}, \bibinfo {author} {\bibfnamefont {V.}~\bibnamefont
  {Tsurkan}}, \bibinfo {author} {\bibfnamefont {A.}~\bibnamefont {Krimmel}},
  \bibinfo {author} {\bibfnamefont {S.}~\bibnamefont {Horn}}, \ and\ \bibinfo
  {author} {\bibfnamefont {A.}~\bibnamefont {Loidl}},\ }\href@noop {}
  {\bibfield  {journal} {\bibinfo  {journal} {J. Phys. Condens. Matter}\
  }\textbf {\bibinfo {volume} {19}},\ \bibinfo {pages} {145262} (\bibinfo
  {year} {2007})}\BibitemShut {NoStop}%
\bibitem [{\citenamefont {Giri}\ \emph {et~al.}(2005)\citenamefont {Giri},
  \citenamefont {Nakamura},\ and\ \citenamefont {Kohara}}]{Giri2005}%
  \BibitemOpen
  \bibfield  {author} {\bibinfo {author} {\bibfnamefont {S.}~\bibnamefont
  {Giri}}, \bibinfo {author} {\bibfnamefont {H.}~\bibnamefont {Nakamura}}, \
  and\ \bibinfo {author} {\bibfnamefont {T.}~\bibnamefont {Kohara}},\
  }\href@noop {} {\bibfield  {journal} {\bibinfo  {journal} {Phys. Rev. B}\
  }\textbf {\bibinfo {volume} {72}},\ \bibinfo {pages} {132404} (\bibinfo
  {year} {2005})}\BibitemShut {NoStop}%
\bibitem [{\citenamefont {B{\"u}ttgen}\ \emph {et~al.}(2006)\citenamefont
  {B{\"u}ttgen}, \citenamefont {Zymara}, \citenamefont {Kegler}, \citenamefont
  {Tsurkan},\ and\ \citenamefont {Loidl}}]{Buettgen2006}%
  \BibitemOpen
  \bibfield  {author} {\bibinfo {author} {\bibfnamefont {N.}~\bibnamefont
  {B{\"u}ttgen}}, \bibinfo {author} {\bibfnamefont {A.}~\bibnamefont {Zymara}},
  \bibinfo {author} {\bibfnamefont {C.}~\bibnamefont {Kegler}}, \bibinfo
  {author} {\bibfnamefont {V.}~\bibnamefont {Tsurkan}}, \ and\ \bibinfo
  {author} {\bibfnamefont {A.}~\bibnamefont {Loidl}},\ }\href@noop {}
  {\bibfield  {journal} {\bibinfo  {journal} {Phys. Rev. B}\ }\textbf {\bibinfo
  {volume} {73}},\ \bibinfo {pages} {132409} (\bibinfo {year}
  {2006})}\BibitemShut {NoStop}%
\bibitem [{\citenamefont {Kalvius}\ \emph {et~al.}(2006)\citenamefont
  {Kalvius}, \citenamefont {Hartmann}, \citenamefont {Krimmel}, \citenamefont
  {Tsurkan},\ and\ \citenamefont {Loidl}}]{Kalvius2006}%
  \BibitemOpen
  \bibfield  {author} {\bibinfo {author} {\bibfnamefont {G.~M.}\ \bibnamefont
  {Kalvius}}, \bibinfo {author} {\bibfnamefont {O.}~\bibnamefont {Hartmann}},
  \bibinfo {author} {\bibfnamefont {A.}~\bibnamefont {Krimmel}}, \bibinfo
  {author} {\bibfnamefont {V.}~\bibnamefont {Tsurkan}}, \ and\ \bibinfo
  {author} {\bibfnamefont {A.}~\bibnamefont {Loidl}},\ }\href@noop {}
  {\bibfield  {journal} {\bibinfo  {journal} {Physica B}\ }\textbf {\bibinfo
  {volume} {378-380}},\ \bibinfo {pages} {592} (\bibinfo {year}
  {2006})}\BibitemShut {NoStop}%
\bibitem [{\citenamefont {Gao}\ \emph {et~al.}(2017)\citenamefont {Gao},
  \citenamefont {Zaharko}, \citenamefont {Tsurkan}, \citenamefont {Su},
  \citenamefont {White}, \citenamefont {Tucker}, \citenamefont {Roessli},
  \citenamefont {Bourdarot}, \citenamefont {Chernyshov}, \citenamefont
  {Fennell}, \citenamefont {Loidl},\ and\ \citenamefont {R{\"u}egg}}]{Gao2017}%
  \BibitemOpen
  \bibfield  {author} {\bibinfo {author} {\bibfnamefont {S.}~\bibnamefont
  {Gao}}, \bibinfo {author} {\bibfnamefont {O.}~\bibnamefont {Zaharko}},
  \bibinfo {author} {\bibfnamefont {V.}~\bibnamefont {Tsurkan}}, \bibinfo
  {author} {\bibfnamefont {Y.}~\bibnamefont {Su}}, \bibinfo {author}
  {\bibfnamefont {J.~S.}\ \bibnamefont {White}}, \bibinfo {author}
  {\bibfnamefont {G.~S.}\ \bibnamefont {Tucker}}, \bibinfo {author}
  {\bibfnamefont {B.}~\bibnamefont {Roessli}}, \bibinfo {author} {\bibfnamefont
  {F.}~\bibnamefont {Bourdarot}}, \bibinfo {author} {\bibfnamefont
  {D.}~\bibnamefont {Chernyshov}}, \bibinfo {author} {\bibfnamefont
  {T.}~\bibnamefont {Fennell}}, \bibinfo {author} {\bibfnamefont
  {A.}~\bibnamefont {Loidl}}, \ and\ \bibinfo {author} {\bibfnamefont
  {C.}~\bibnamefont {R{\"u}egg}},\ }\href@noop {} {\bibfield  {journal}
  {\bibinfo  {journal} {Nat. Phys.}\ }\textbf {\bibinfo {volume} {13}},\
  \bibinfo {pages} {157} (\bibinfo {year} {2017})}\BibitemShut {NoStop}%
\bibitem [{\citenamefont {MacDougall}\ \emph {et~al.}(2016)\citenamefont
  {MacDougall}, \citenamefont {Aczel}, \citenamefont {Su}, \citenamefont
  {Schweika}, \citenamefont {Faulhaber}, \citenamefont {Schneidewind},
  \citenamefont {Christianson}, \citenamefont {Zaretsky}, \citenamefont {Zhou},
  \citenamefont {Mandrus},\ and\ \citenamefont {Nagler}}]{GJM2016}%
  \BibitemOpen
  \bibfield  {author} {\bibinfo {author} {\bibfnamefont {G.~J.}\ \bibnamefont
  {MacDougall}}, \bibinfo {author} {\bibfnamefont {A.~A.}\ \bibnamefont
  {Aczel}}, \bibinfo {author} {\bibfnamefont {Y.}~\bibnamefont {Su}}, \bibinfo
  {author} {\bibfnamefont {W.}~\bibnamefont {Schweika}}, \bibinfo {author}
  {\bibfnamefont {E.}~\bibnamefont {Faulhaber}}, \bibinfo {author}
  {\bibfnamefont {A.}~\bibnamefont {Schneidewind}}, \bibinfo {author}
  {\bibfnamefont {A.~D.}\ \bibnamefont {Christianson}}, \bibinfo {author}
  {\bibfnamefont {J.~L.}\ \bibnamefont {Zaretsky}}, \bibinfo {author}
  {\bibfnamefont {H.~D.}\ \bibnamefont {Zhou}}, \bibinfo {author}
  {\bibfnamefont {D.}~\bibnamefont {Mandrus}}, \ and\ \bibinfo {author}
  {\bibfnamefont {S.~E.}\ \bibnamefont {Nagler}},\ }\href@noop {} {\bibfield
  {journal} {\bibinfo  {journal} {Phys. Rev. B}\ }\textbf {\bibinfo {volume}
  {94}},\ \bibinfo {pages} {184422} (\bibinfo {year} {2016})}\BibitemShut
  {NoStop}%
\bibitem [{\citenamefont {Roy}\ \emph {et~al.}(2013)\citenamefont {Roy},
  \citenamefont {Pandey}, \citenamefont {Zhang}, \citenamefont {Heitmann},
  \citenamefont {Vaknir}, \citenamefont {Johnston},\ and\ \citenamefont
  {Furukawa}}]{Roy2013}%
  \BibitemOpen
  \bibfield  {author} {\bibinfo {author} {\bibfnamefont {B.}~\bibnamefont
  {Roy}}, \bibinfo {author} {\bibfnamefont {A.}~\bibnamefont {Pandey}},
  \bibinfo {author} {\bibfnamefont {Q.}~\bibnamefont {Zhang}}, \bibinfo
  {author} {\bibfnamefont {T.~W.}\ \bibnamefont {Heitmann}}, \bibinfo {author}
  {\bibfnamefont {D.}~\bibnamefont {Vaknir}}, \bibinfo {author} {\bibfnamefont
  {D.~C.}\ \bibnamefont {Johnston}}, \ and\ \bibinfo {author} {\bibfnamefont
  {Y.}~\bibnamefont {Furukawa}},\ }\href@noop {} {\bibfield  {journal}
  {\bibinfo  {journal} {Phys. Rev. B}\ }\textbf {\bibinfo {volume} {88}},\
  \bibinfo {pages} {174415} (\bibinfo {year} {2013})}\BibitemShut {NoStop}%
\bibitem [{\citenamefont {Ge}\ \emph {et~al.}(2017)\citenamefont {Ge},
  \citenamefont {Paddison}, \citenamefont {Stone}, \citenamefont {Calder},
  \citenamefont {Subramanian}, \citenamefont {Ramirez},\ and\ \citenamefont
  {Mourigal}}]{Mourigal2017}%
  \BibitemOpen
  \bibfield  {author} {\bibinfo {author} {\bibfnamefont {L.}~\bibnamefont
  {Ge}}, \bibinfo {author} {\bibfnamefont {J.~A.~M.}\ \bibnamefont {Paddison}},
  \bibinfo {author} {\bibfnamefont {M.~B.}\ \bibnamefont {Stone}}, \bibinfo
  {author} {\bibfnamefont {S.}~\bibnamefont {Calder}}, \bibinfo {author}
  {\bibfnamefont {M.~A.}\ \bibnamefont {Subramanian}}, \bibinfo {author}
  {\bibfnamefont {A.~P.}\ \bibnamefont {Ramirez}}, \ and\ \bibinfo {author}
  {\bibfnamefont {M.}~\bibnamefont {Mourigal}},\ }\href@noop {} {\bibfield
  {journal} {\bibinfo  {journal} {Phys. Rev. B}\ } (\bibinfo {year}
  {2017})}\BibitemShut {NoStop}%
\bibitem [{\citenamefont {Chamorro}\ \emph {et~al.}(2018)\citenamefont
  {Chamorro}, \citenamefont {Ge}, \citenamefont {Flynn}, \citenamefont
  {Subramanian}, \citenamefont {Mourigal},\ and\ \citenamefont
  {McQueen}}]{chamorro2018}%
  \BibitemOpen
  \bibfield  {author} {\bibinfo {author} {\bibfnamefont {J.~R.}\ \bibnamefont
  {Chamorro}}, \bibinfo {author} {\bibfnamefont {L.}~\bibnamefont {Ge}},
  \bibinfo {author} {\bibfnamefont {J.}~\bibnamefont {Flynn}}, \bibinfo
  {author} {\bibfnamefont {M.~A.}\ \bibnamefont {Subramanian}}, \bibinfo
  {author} {\bibfnamefont {M.}~\bibnamefont {Mourigal}}, \ and\ \bibinfo
  {author} {\bibfnamefont {T.~M.}\ \bibnamefont {McQueen}},\ }\href {\doibase
  10.1103/PhysRevMaterials.2.034404} {\bibfield  {journal} {\bibinfo  {journal}
  {Phys. Rev. Materials}\ }\textbf {\bibinfo {volume} {2}},\ \bibinfo {pages}
  {034404} (\bibinfo {year} {2018})}\BibitemShut {NoStop}%
\bibitem [{\citenamefont {Ismunandar}\ \emph {et~al.}(1999)\citenamefont
  {Ismunandar}, \citenamefont {Kennedy},\ and\ \citenamefont
  {Hunter}}]{Kennedy99}%
  \BibitemOpen
  \bibfield  {author} {\bibinfo {author} {\bibnamefont {Ismunandar}}, \bibinfo
  {author} {\bibfnamefont {B.~J.}\ \bibnamefont {Kennedy}}, \ and\ \bibinfo
  {author} {\bibfnamefont {B.~A.}\ \bibnamefont {Hunter}},\ }\href@noop {}
  {\bibfield  {journal} {\bibinfo  {journal} {Mat. Res. Bull.}\ }\textbf
  {\bibinfo {volume} {34}},\ \bibinfo {pages} {135} (\bibinfo {year}
  {1999})}\BibitemShut {NoStop}%
\bibitem [{\citenamefont {Dollase}\ and\ \citenamefont
  {O'Neill}(1997)}]{Dollase97}%
  \BibitemOpen
  \bibfield  {author} {\bibinfo {author} {\bibfnamefont {W.~A.}\ \bibnamefont
  {Dollase}}\ and\ \bibinfo {author} {\bibfnamefont {H.~S.~C.}\ \bibnamefont
  {O'Neill}},\ }\href@noop {} {\bibfield  {journal} {\bibinfo  {journal} {Acta
  Cryst.}\ }\textbf {\bibinfo {volume} {C53}},\ \bibinfo {pages} {657}
  (\bibinfo {year} {1997})}\BibitemShut {NoStop}%
\bibitem [{\citenamefont {Endoh}\ \emph {et~al.}(1999)\citenamefont {Endoh},
  \citenamefont {Fujishima}, \citenamefont {Atake}, \citenamefont {Matsumoto},
  \citenamefont {Hayashi},\ and\ \citenamefont {Nagata}}]{Endoh99}%
  \BibitemOpen
  \bibfield  {author} {\bibinfo {author} {\bibfnamefont {R.}~\bibnamefont
  {Endoh}}, \bibinfo {author} {\bibfnamefont {O.}~\bibnamefont {Fujishima}},
  \bibinfo {author} {\bibfnamefont {T.}~\bibnamefont {Atake}}, \bibinfo
  {author} {\bibfnamefont {N.}~\bibnamefont {Matsumoto}}, \bibinfo {author}
  {\bibfnamefont {M.}~\bibnamefont {Hayashi}}, \ and\ \bibinfo {author}
  {\bibfnamefont {S.}~\bibnamefont {Nagata}},\ }\href@noop {} {\bibfield
  {journal} {\bibinfo  {journal} {J. Phys. Chem. Sol.}\ }\textbf {\bibinfo
  {volume} {60}},\ \bibinfo {pages} {457} (\bibinfo {year} {1999})}\BibitemShut
  {NoStop}%
\bibitem [{\citenamefont {Naka}\ \emph {et~al.}(2015)\citenamefont {Naka},
  \citenamefont {Sato}, \citenamefont {Matsushita}, \citenamefont {Terada},
  \citenamefont {Ishii}, \citenamefont {Nakane}, \citenamefont {Taguchi},\ and\
  \citenamefont {Matsushita}}]{Naka2015}%
  \BibitemOpen
  \bibfield  {author} {\bibinfo {author} {\bibfnamefont {T.}~\bibnamefont
  {Naka}}, \bibinfo {author} {\bibfnamefont {K.}~\bibnamefont {Sato}}, \bibinfo
  {author} {\bibfnamefont {Y.}~\bibnamefont {Matsushita}}, \bibinfo {author}
  {\bibfnamefont {N.}~\bibnamefont {Terada}}, \bibinfo {author} {\bibfnamefont
  {S.}~\bibnamefont {Ishii}}, \bibinfo {author} {\bibfnamefont
  {T.}~\bibnamefont {Nakane}}, \bibinfo {author} {\bibfnamefont
  {M.}~\bibnamefont {Taguchi}}, \ and\ \bibinfo {author} {\bibfnamefont
  {A.}~\bibnamefont {Matsushita}},\ }\href@noop {} {\bibfield  {journal}
  {\bibinfo  {journal} {Phys. Rev. B}\ }\textbf {\bibinfo {volume} {91}},\
  \bibinfo {pages} {224412} (\bibinfo {year} {2015})}\BibitemShut {NoStop}%
\bibitem [{\citenamefont {Melot}\ \emph {et~al.}(2009)\citenamefont {Melot},
  \citenamefont {Page}, \citenamefont {Seshadri}, \citenamefont {Balents},\
  and\ \citenamefont {Bergman}}]{Melot2009}%
  \BibitemOpen
  \bibfield  {author} {\bibinfo {author} {\bibfnamefont {B.~C.}\ \bibnamefont
  {Melot}}, \bibinfo {author} {\bibfnamefont {K.}~\bibnamefont {Page}},
  \bibinfo {author} {\bibfnamefont {R.}~\bibnamefont {Seshadri}}, \bibinfo
  {author} {\bibfnamefont {L.}~\bibnamefont {Balents}}, \ and\ \bibinfo
  {author} {\bibfnamefont {D.}~\bibnamefont {Bergman}},\ }\href@noop {}
  {\bibfield  {journal} {\bibinfo  {journal} {Phys. Rev. B}\ }\textbf {\bibinfo
  {volume} {80}},\ \bibinfo {pages} {104420} (\bibinfo {year}
  {2009})}\BibitemShut {NoStop}%
\bibitem [{\citenamefont {Bitoh}\ \emph {et~al.}(1992)\citenamefont {Bitoh},
  \citenamefont {Hagino}, \citenamefont {Seki}, \citenamefont {Chikazawa},\
  and\ \citenamefont {Nagata}}]{Bitoh92}%
  \BibitemOpen
  \bibfield  {author} {\bibinfo {author} {\bibfnamefont {T.}~\bibnamefont
  {Bitoh}}, \bibinfo {author} {\bibfnamefont {T.}~\bibnamefont {Hagino}},
  \bibinfo {author} {\bibfnamefont {Y.}~\bibnamefont {Seki}}, \bibinfo {author}
  {\bibfnamefont {S.}~\bibnamefont {Chikazawa}}, \ and\ \bibinfo {author}
  {\bibfnamefont {S.}~\bibnamefont {Nagata}},\ }\href@noop {} {\bibfield
  {journal} {\bibinfo  {journal} {J. Phys. Soc. Jpn.}\ }\textbf {\bibinfo
  {volume} {61}},\ \bibinfo {pages} {3011} (\bibinfo {year}
  {1992})}\BibitemShut {NoStop}%
\bibitem [{\citenamefont {Hagino}\ \emph {et~al.}(1995)\citenamefont {Hagino},
  \citenamefont {Seki}, \citenamefont {Wada}, \citenamefont {Tsuji},
  \citenamefont {Shirane}, \citenamefont {Kumagai},\ and\ \citenamefont
  {Nagata}}]{Hagino95}%
  \BibitemOpen
  \bibfield  {author} {\bibinfo {author} {\bibfnamefont {T.}~\bibnamefont
  {Hagino}}, \bibinfo {author} {\bibfnamefont {Y.}~\bibnamefont {Seki}},
  \bibinfo {author} {\bibfnamefont {N.}~\bibnamefont {Wada}}, \bibinfo {author}
  {\bibfnamefont {S.}~\bibnamefont {Tsuji}}, \bibinfo {author} {\bibfnamefont
  {T.}~\bibnamefont {Shirane}}, \bibinfo {author} {\bibfnamefont
  {K.}~\bibnamefont {Kumagai}}, \ and\ \bibinfo {author} {\bibfnamefont
  {S.}~\bibnamefont {Nagata}},\ }\href@noop {} {\bibfield  {journal} {\bibinfo
  {journal} {Phys. Rev. B}\ }\textbf {\bibinfo {volume} {51}},\ \bibinfo
  {pages} {12673} (\bibinfo {year} {1995})}\BibitemShut {NoStop}%
\bibitem [{\citenamefont {Arlett}(1968)}]{Arlett68}%
  \BibitemOpen
  \bibfield  {author} {\bibinfo {author} {\bibfnamefont {R.~H.}\ \bibnamefont
  {Arlett}},\ }\href@noop {} {\bibfield  {journal} {\bibinfo  {journal} {J. Am.
  Cer. Soc.}\ }\textbf {\bibinfo {volume} {51}},\ \bibinfo {pages} {292}
  (\bibinfo {year} {1968})}\BibitemShut {NoStop}%
\bibitem [{\citenamefont {Mizoguchi}\ \emph {et~al.}(2002)\citenamefont
  {Mizoguchi}, \citenamefont {Hirano}, \citenamefont {Fujitsu}, \citenamefont
  {Takeuchi}, \citenamefont {Ueda},\ and\ \citenamefont
  {Hosono}}]{Mizoguchi2002}%
  \BibitemOpen
  \bibfield  {author} {\bibinfo {author} {\bibfnamefont {H.}~\bibnamefont
  {Mizoguchi}}, \bibinfo {author} {\bibfnamefont {M.}~\bibnamefont {Hirano}},
  \bibinfo {author} {\bibfnamefont {S.}~\bibnamefont {Fujitsu}}, \bibinfo
  {author} {\bibfnamefont {T.}~\bibnamefont {Takeuchi}}, \bibinfo {author}
  {\bibfnamefont {K.}~\bibnamefont {Ueda}}, \ and\ \bibinfo {author}
  {\bibfnamefont {H.}~\bibnamefont {Hosono}},\ }\href@noop {} {\bibfield
  {journal} {\bibinfo  {journal} {Appl. Phys. Lett.}\ }\textbf {\bibinfo
  {volume} {80}},\ \bibinfo {pages} {1207} (\bibinfo {year}
  {2002})}\BibitemShut {NoStop}%
\bibitem [{\citenamefont {Mansourian-Hadavi}\ \emph {et~al.}(2010)\citenamefont
  {Mansourian-Hadavi}, \citenamefont {Wansom}, \citenamefont {Perry},
  \citenamefont {Nagaraja}, \citenamefont {Mason}, \citenamefont {Ye},\ and\
  \citenamefont {Freeman}}]{MH2010}%
  \BibitemOpen
  \bibfield  {author} {\bibinfo {author} {\bibfnamefont {N.}~\bibnamefont
  {Mansourian-Hadavi}}, \bibinfo {author} {\bibfnamefont {S.}~\bibnamefont
  {Wansom}}, \bibinfo {author} {\bibfnamefont {N.~H.}\ \bibnamefont {Perry}},
  \bibinfo {author} {\bibfnamefont {A.~R.}\ \bibnamefont {Nagaraja}}, \bibinfo
  {author} {\bibfnamefont {T.~O.}\ \bibnamefont {Mason}}, \bibinfo {author}
  {\bibfnamefont {L.}~\bibnamefont {Ye}}, \ and\ \bibinfo {author}
  {\bibfnamefont {A.~J.}\ \bibnamefont {Freeman}},\ }\href@noop {} {\bibfield
  {journal} {\bibinfo  {journal} {Phys. Rev. B}\ }\textbf {\bibinfo {volume}
  {81}},\ \bibinfo {pages} {075112} (\bibinfo {year} {2010})}\BibitemShut
  {NoStop}%
\bibitem [{\citenamefont {Garlea}\ \emph {et~al.}(2010)\citenamefont {Garlea},
  \citenamefont {Chakoumakos}, \citenamefont {Moore}, \citenamefont {Taylor},
  \citenamefont {Chae}, \citenamefont {Maples}, \citenamefont {Riedel},
  \citenamefont {Lynn},\ and\ \citenamefont {Selby}}]{Garlea2010}%
  \BibitemOpen
  \bibfield  {author} {\bibinfo {author} {\bibfnamefont {V.~O.}\ \bibnamefont
  {Garlea}}, \bibinfo {author} {\bibfnamefont {B.~C.}\ \bibnamefont
  {Chakoumakos}}, \bibinfo {author} {\bibfnamefont {S.~A.}\ \bibnamefont
  {Moore}}, \bibinfo {author} {\bibfnamefont {G.~B.}\ \bibnamefont {Taylor}},
  \bibinfo {author} {\bibfnamefont {T.}~\bibnamefont {Chae}}, \bibinfo {author}
  {\bibfnamefont {R.~G.}\ \bibnamefont {Maples}}, \bibinfo {author}
  {\bibfnamefont {R.~A.}\ \bibnamefont {Riedel}}, \bibinfo {author}
  {\bibfnamefont {G.~W.}\ \bibnamefont {Lynn}}, \ and\ \bibinfo {author}
  {\bibfnamefont {D.~L.}\ \bibnamefont {Selby}},\ }\href@noop {} {\bibfield
  {journal} {\bibinfo  {journal} {Applied Physics A}\ }\textbf {\bibinfo
  {volume} {99}},\ \bibinfo {pages} {531} (\bibinfo {year} {2010})}\BibitemShut
  {NoStop}%
\bibitem [{\citenamefont {Rodriguez-Carvajal}(1993)}]{FULLPROF93}%
  \BibitemOpen
  \bibfield  {author} {\bibinfo {author} {\bibfnamefont {J.}~\bibnamefont
  {Rodriguez-Carvajal}},\ }\href@noop {} {\bibfield  {journal} {\bibinfo
  {journal} {Physica B}\ }\textbf {\bibinfo {volume} {192}},\ \bibinfo {pages}
  {55} (\bibinfo {year} {1993})}\BibitemShut {NoStop}%
\bibitem [{\citenamefont {Hanashima}\ \emph {et~al.}(2013)\citenamefont
  {Hanashima}, \citenamefont {Kodama}, \citenamefont {Akahoshi}, \citenamefont
  {Kanadani},\ and\ \citenamefont {Saito}}]{Hanashima2013}%
  \BibitemOpen
  \bibfield  {author} {\bibinfo {author} {\bibfnamefont {K.}~\bibnamefont
  {Hanashima}}, \bibinfo {author} {\bibfnamefont {Y.}~\bibnamefont {Kodama}},
  \bibinfo {author} {\bibfnamefont {D.}~\bibnamefont {Akahoshi}}, \bibinfo
  {author} {\bibfnamefont {C.}~\bibnamefont {Kanadani}}, \ and\ \bibinfo
  {author} {\bibfnamefont {T.}~\bibnamefont {Saito}},\ }\href@noop {}
  {\bibfield  {journal} {\bibinfo  {journal} {J. Phys. Soc. Jpn.}\ }\textbf
  {\bibinfo {volume} {82}},\ \bibinfo {pages} {024702} (\bibinfo {year}
  {2013})}\BibitemShut {NoStop}%
\bibitem [{\citenamefont {Scholl}\ and\ \citenamefont
  {Binder}(1980)}]{Scholl80}%
  \BibitemOpen
  \bibfield  {author} {\bibinfo {author} {\bibfnamefont {F.}~\bibnamefont
  {Scholl}}\ and\ \bibinfo {author} {\bibfnamefont {K.}~\bibnamefont
  {Binder}},\ }\href@noop {} {\bibfield  {journal} {\bibinfo  {journal} {Z.
  Phys. B}\ }\textbf {\bibinfo {volume} {39}},\ \bibinfo {pages} {239}
  (\bibinfo {year} {1980})}\BibitemShut {NoStop}%
\bibitem [{\citenamefont {Blasse}(1963)}]{Blasse61}%
  \BibitemOpen
  \bibfield  {author} {\bibinfo {author} {\bibfnamefont {G.}~\bibnamefont
  {Blasse}},\ }\href@noop {} {\bibfield  {journal} {\bibinfo  {journal}
  {Phillips Res. Repts}\ }\textbf {\bibinfo {volume} {18}},\ \bibinfo {pages}
  {383} (\bibinfo {year} {1963})}\BibitemShut {NoStop}%
\bibitem [{\citenamefont {Bertaut}\ \emph {et~al.}(1959)\citenamefont
  {Bertaut}, \citenamefont {Forrat},\ and\ \citenamefont {Dulac}}]{Bertaut59}%
  \BibitemOpen
  \bibfield  {author} {\bibinfo {author} {\bibfnamefont {F.}~\bibnamefont
  {Bertaut}}, \bibinfo {author} {\bibfnamefont {F.}~\bibnamefont {Forrat}}, \
  and\ \bibinfo {author} {\bibfnamefont {J.}~\bibnamefont {Dulac}},\
  }\href@noop {} {\bibfield  {journal} {\bibinfo  {journal} {C. A. Acad. Sci.
  Paris (Compt. rend.)}\ }\textbf {\bibinfo {volume} {249}},\ \bibinfo {pages}
  {726} (\bibinfo {year} {1959})}\BibitemShut {NoStop}%
\end{thebibliography}
\providecommand{\noopsort}[1]{}\providecommand{\singleletter}[1]{#1}%

\end{document}